\documentclass[sigconf]{acmart}
\usepackage{graphicx}
\usepackage{lipsum}
\usepackage{url}
\graphicspath{{Figures/}}
\usepackage{caption}
\usepackage{subfig} 
\usepackage{color}
\usepackage{balance}

\usepackage{booktabs} % For formal tables

\setcopyright{acmcopyright}
\acmDOI{10.1145/3313991.3314006}

\acmISBN{978-1-4503-6287-0/19/02}

\acmConference[ICCAE 2019]{11th International Conference on Computer and Automation Engineering}{February 23-25, 2019}{Perth, WN, Australia}
\begin{document}
\title{TEST: A Terminology Extraction System for \\Technology Related Terms}

\author{Murhaf Hossari}
\affiliation{%
  \institution{ADAPT Centre, Trinity College Dublin}
  \streetaddress{ }
  \city{Dublin}
  \state{Ireland}
  \postcode{ }
}
\email{murhaf.hossari@adaptcentre.ie}

\author{Soumyabrata Dev}
\affiliation{%
  \institution{ADAPT Centre, Trinity College Dublin}
  \streetaddress{ }
  \city{Dublin}
  \state{Ireland}
  \postcode{ }
}
\email{soumyabrata.dev@adaptcentre.ie}

\author{John D.\ Kelleher}
\affiliation{%
  \institution{ADAPT Centre, Technological University Dublin}
  \streetaddress{ }
  \city{Dublin}
  \country{Ireland}}
\email{john.d.kelleher@dit.ie}

\renewcommand{\shortauthors}{M. Hossari et al.}

% To remove ACM reference format
%\settopmatter{printacmref=false}

\begin{abstract}
Tracking developments in the highly dynamic data-technology landscape are vital to keeping up with novel technologies and tools, in the various areas of Artificial Intelligence (AI). However, It is difficult to keep track of all the relevant technology keywords. In this paper, we propose a novel system that addresses this problem. This tool is used to automatically detect the existence of new technologies and tools in text, and extract terms used to describe these new technologies. The extracted new terms can be logged as new AI technologies as they are found on-the-fly in the web. It can be subsequently classified into the relevant semantic labels and AI domains. Our proposed tool is based on a two-stage cascading model -- the first stage classifies if the sentence contains a technology term or not; and the second stage identifies the technology keyword in the sentence. We obtain a competitive accuracy for both tasks of sentence classification and text identification.
\end{abstract}




\maketitle

\section{Introduction}
With the ubiquity of online resources and the massive growth of news and blog articles, it has been extremely difficult for the end users to keep abreast of all the latest advances. Nowadays, data science and artificial intelligence have been applied to a multitude of applications~\cite{kelleher2018data}, ranging from e-mail spam filtering, advertising and marketing industry~\cite{nautiyal2018advert, hossari2018ADNet}, and social media. 
Particularly, in these realm of online articles and blog articles, the growth in the amount of information is almost in an exponential nature. Therefore, it is important for the researchers to develop a system that can automatically parse the millions of documents, and identify the key technological terms. Such system will greatly reduce the manual work, and save several man-hours. 

In this paper, we develop a machine-learning based solution, that is capable of the discovery and extraction of information related to AI technologies. We call the system \textbf{TEST}, that stands for \textbf{T}erminology \textbf{E}xtraction \textbf{S}ystem for \textbf{T}echnology-related terms. This will greatly help the users to build the knowledge about these technologies, and also predict the trends and sentiments around them by observing the textual data found in tech news articles and blogs.

\begin{figure*}[htb]
\centering
\includegraphics[trim={3cm 7.5cm 0 0}, width=0.8\textwidth]{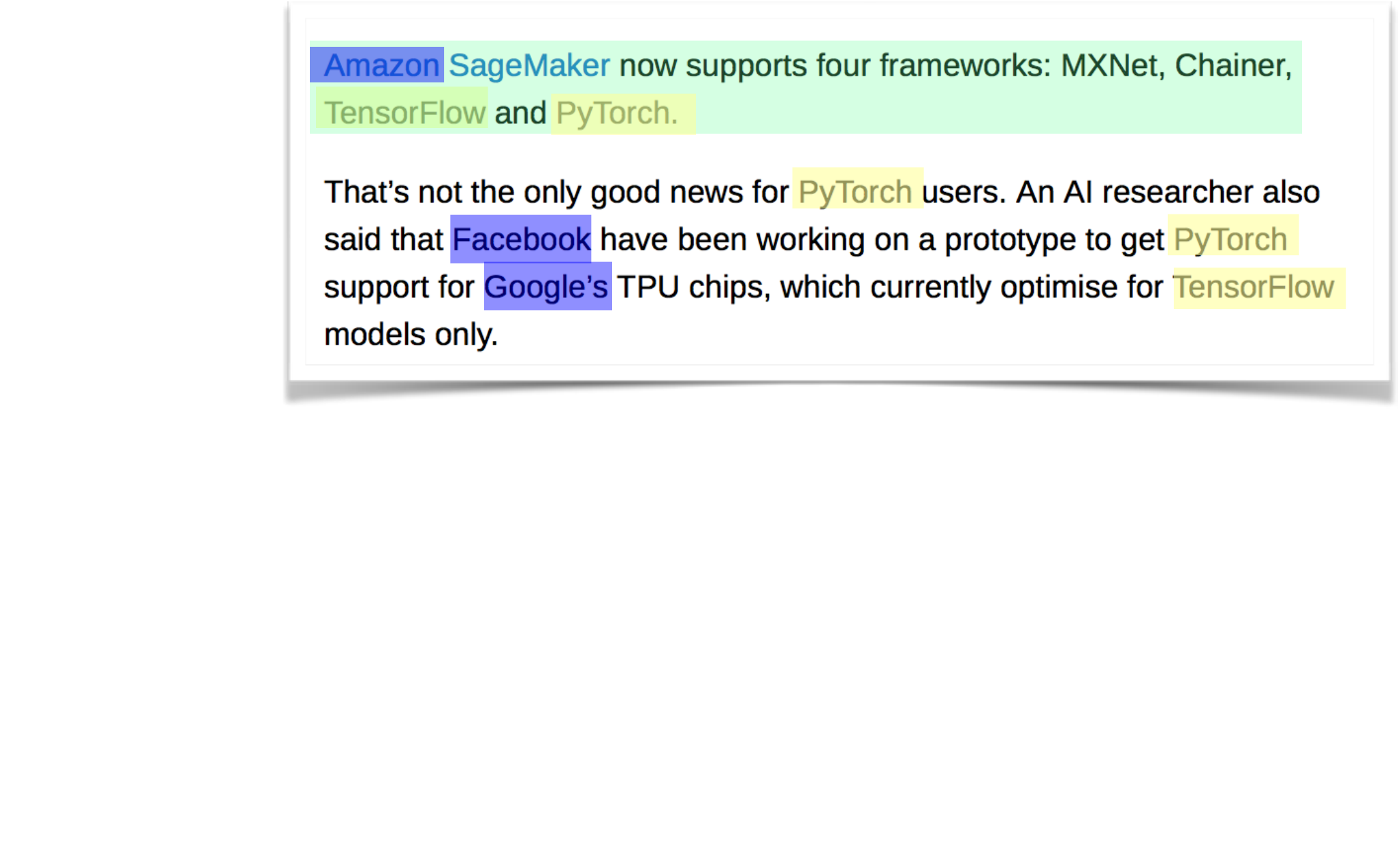}
\caption{Illustration of TEST system that automatically identifies the tech keywords from a sentence. The sentence containing technology term are highlighted in green colour. The extracted words are also colour-coded in terms of their semantics. Technology keywords are marked in yellow, and organisation names are marked in blue colour. (Text snippet extracted from \url{theregister.co.uk}.)}
\label{fig:test-system}
\end{figure*}

To achieve this, we combined AI techniques such as supervised and unsupervised learning on our text datasets. We use the concepts drawn from natural language processing, such as word embeddings, and conventional machine learning techniques to propose a state-of-the-art text extraction system. Our proposed ensemble method achieved a competitive performance for both sentence classification and term classification stages. The system will help the users to get insights about AI technologies in the market, and how they interact or relate to each other. This would have a great impact on decision making, and would reduce the manual review needed to peruse daily news articles to be able to analyse the top trends. This solution will make the analysis task more focused and clear.

The extracted data and insights can be also used and augmented to build more visually helpful insights. This can eventually be used to create short article snippets, that would summarise the news articles and present structured and meaningful insights about the data science technologies, in a clear and concise fashion.

\section{Related Work}

In the literature, there are plethora of application areas for text classification. It is extensively used in news filtering and spam detection~\cite{sahami1998bayesian}, document categorisation and text summarisation~\cite{chakrabarti1997using}. A good survey of text classification algorithms can be found in ~\cite{aggarwal2012survey}. Most of the traditional techniques use decision trees, rule-based classifiers, bayesian classifiers etc. Recently, neural-network based deep networks are also used for the purpose of text summarisation and text classification. However, they do not propose a working system for these purposes -- we bridge this gap in the literature by proposing a novel technology extraction system.

Figure~\ref{fig:test-system} illustrates our proposed TEST system. As an illustration, we copy verbatim a text paragraph from \url{theregister.co.uk}. Our TEST system parses each sentence in the paragraph. It identifies whether a sentence in the paragraph contains a technology-related term. In the next stage of the cascading model, the TEST system automatically identifies the keywords in the sentence. In our example, the keywords \emph{TensorFlow} and \emph{PyTorch} are identified as the technology terms, while \emph{Google} can be identified as an organisation name and linked at a later stage to the technology \emph{TensorFlow} as having a ownership relation. 

The rest of the paper is organised as follows. Section~\ref{sec:dataset} discusses the method, in which we prepare the dataset from internet articles and tech blogs. In Section~\ref{sec:cascade}, we discuss the two-stage cascading model of our proposed TEST system. We discuss the objective evaluation of our proposed system in Section~\ref{sec:results}. Finally, Section~\ref{sec:conclusion} concludes the paper, and discusses the potential future work.

\section{Dataset preparation}
\label{sec:dataset}

In order to train a machine learning model that is capable of extracting technology terms from text, we need a labelled dataset where the technology terms are properly annotated and tagged. To the best of our knowledge, there are no datasets available in the literature that classifies whether a sentence contains a technological keyword, and also annotates the technological keyword in a sentence. 
Therefore, we created a new dataset of example texts, that mostly comes from technology news articles and blogs. We crawled approximately $500$ articles from various online sources, that are varied in size. We created a dataset of $150$ thousands paragraphs. Together with this dataset, we also manually populated a list of technology keywords comprising single and multi-word technology terms. For example, the list contains single technology terms such as \emph{Cortana} and, multiple-word technology terms as \emph{Apache Hive} and \emph{Google Cloud Natural Language API}.

Using this manually collected list of technology terms, we attempted to automatically annotate the crawled text articles. We used string matching, and we annotated every token in the text as being part of technology term \texttt{T} or other \texttt{O}. The result of this process was $3310000$ tokens. The resulting dataset was highly imbalanced, due to the fact that technology terms are less frequent than the rest of the vocabulary. In order to reduce the imbalance, we propose a cascading model where the tokens are grouped into bigger blocks (for example sentences), and then the whole block is labelled based on the fact that it contains a technology term or not. More details of this cascading model is described in the subsequent section. We used this dataset of $3.3$ million tokens for our subsequent experiments.

\section{Cascading method}
\label{sec:cascade}
In this section, we propose a cascading method that is useful in our scenario of unbalanced dataset. The first stage of the cascading method classifies whether a sentence contains a technology term or not. In case the sentence contains a technology term, the second stage of the cascading method is used, which extracts the technology keyword from the sentence. Such cascading method greatly helps in ignoring the cases when the sentence does not contain a technology term. 
The cascading method looks at sentence level classification as a first step, and then looks at term extraction. For this purpose, we group tokens in larger groups/collections for example sentences, and we tag each sentence as `contains a technology term' or `doesn't contain a technology term'. 

We run the automatic annotation using the previously mentioned pattern of string matching. The resulting dataset contains $243336$ sentences, out of which only about $10$ thousands sentences were positive examples (i.e.\ contain technology terms). This resulting dataset is highly unbalanced, as there are more instances of sentences containing a technology term, as compared to otherwise. We use a random downsampling technique to remove the impact of imbalance nature, and generate a balanced dataset. The positive samples are the minority cases, whereas the negative samples are larger in number and constitute the majority cases. In order to create a balanced dataset, we consider all the $10$ thousand positive examples, and then perform a random selection of $10$ thousand negative examples from the remaining sentences. The resultant dataset is balanced in nature, containing equal number of positive- and negative- samples. We use this balanced dataset for our subsequent experiments and analysis.

\subsection{Text Classification}

In our proposed system, we use Facebook's fastText~\cite{joulin2016bag} implementation to train our own word representations using all the textual corpus we have. We used the \texttt{skipgram} method in order to learn the word embeddings which is based on the concepts drawn from deep learning for natural language processing, and particularly \texttt{word2vec} method. It involves representing words as n-dimensional vectors. Subsequently, we used these word vectors to represent the sentences. We train our model using the corpus of $243336$ sentences using the mentioned approach. After training, each word in the sentence is represented using a $300$ dimensional vector. We average the vector representations of all the words in a sentence, across their corresponding elements to compute the vector representation of the entire sentence. Using this approach, we represent each sentence in the corpus with a $300$ dimensional vector also.

Finally, we use a softmax function $f$ over the binary labels -- \texttt{T} and \texttt{O} to estimate the final label of the sentence. We assume that the total number of sentences in the corpus is $N$. Our objective~\cite{joulin2016bag} in this task of text classification is to minimise the following objective function:

\begin{equation}
\label{eq:smax}
-\frac{1}{N}\sum_{n=1}^{N} y_{n}log(f(BAx_{n})),
\end{equation}

where $A$, $B$ are weight matrices, $x_n$ in the feature vector of $n$-th sentence, and $y_n$ is the corresponding binary label.

\subsection{Term Extraction}
In the second stage of the cascading model, we are interested in extracting the tech keyword from the sentence. 
Similar to the text classification stage, the term extraction stage also considers the same dataset of $10$ thousand positive examples, wherein each example/sentence contains a technology term. Using this dataset of $10$ thousand positive samples, we expanded into a more detailed dataset comprising the individual tokens of the sentences. We tagged each token in the $10$ thousand sentences individually, and labelled each token with a binary label -- token is part of a technology term labelled as \texttt{T}, and token is not part of a technology term labelled as \texttt{O}. This newly created dataset is used for the second layer of the cascading model, which deals with term extraction.

We use the labelled tokens in the sentences to train the term extraction model using Stanford Named Entity Recognition (NER) tool~\cite{finkel2005incorporating}. The NER tool uses well-engineered natural language processing feature descriptors, and represents each sentences of the corpus into discriminative features. We train a Conditional Random Field (CRF) sequence model, using the generated features. The output of the CRF model labels each token in a sentence as \texttt{T} or \texttt{O}. Hence, we can identify the technological term(s) in a single sentence as a sequence of tokens with \texttt{T} labels.

We use the following features to train the CRF model:

\begin{itemize}
\item Current Word
\item Previous Word
\item Next Word
\item Current Word Character all n-grams
\item Current POS Tag
\item Surrounding POS Tag Sequence
\item Current Word Shape
\item Surrounding Word Shape Sequence
\item Presence of Word in Left Window (window size $4$)
\item Presence of Word in Right Window (window size $4$)
\end{itemize}

The CRF model for NER is a well-established model, in order to estimate the probability of a hidden state~\footnote{In the areas of natural language processing, a state is defined as one of the possible events, that constitutes the stochastic model of the CRF.}, with some \emph{a priori} given observations~\cite{finkel2005incorporating}. We define the transition probabilities between two adjacent states as $\phi_c$. The term $\phi_c$ is often described as the \emph{clique potential}. Therefore, we can define the probability $p$ of a chain of state sequences as:

\begin{equation}
\label{eq:clik}
p = \prod_{i=1}^{N} \phi_i (s_{i-1}, s_i),
\end{equation}

where $\phi_i$ is the clique potential for position $i$, with respect to the transition between states $s_{i-1}$ and $s_{i}$.

\begin{figure}[htb]
\centering
\includegraphics[trim={0.7cm 4cm 0 0}, width=0.42\textwidth]{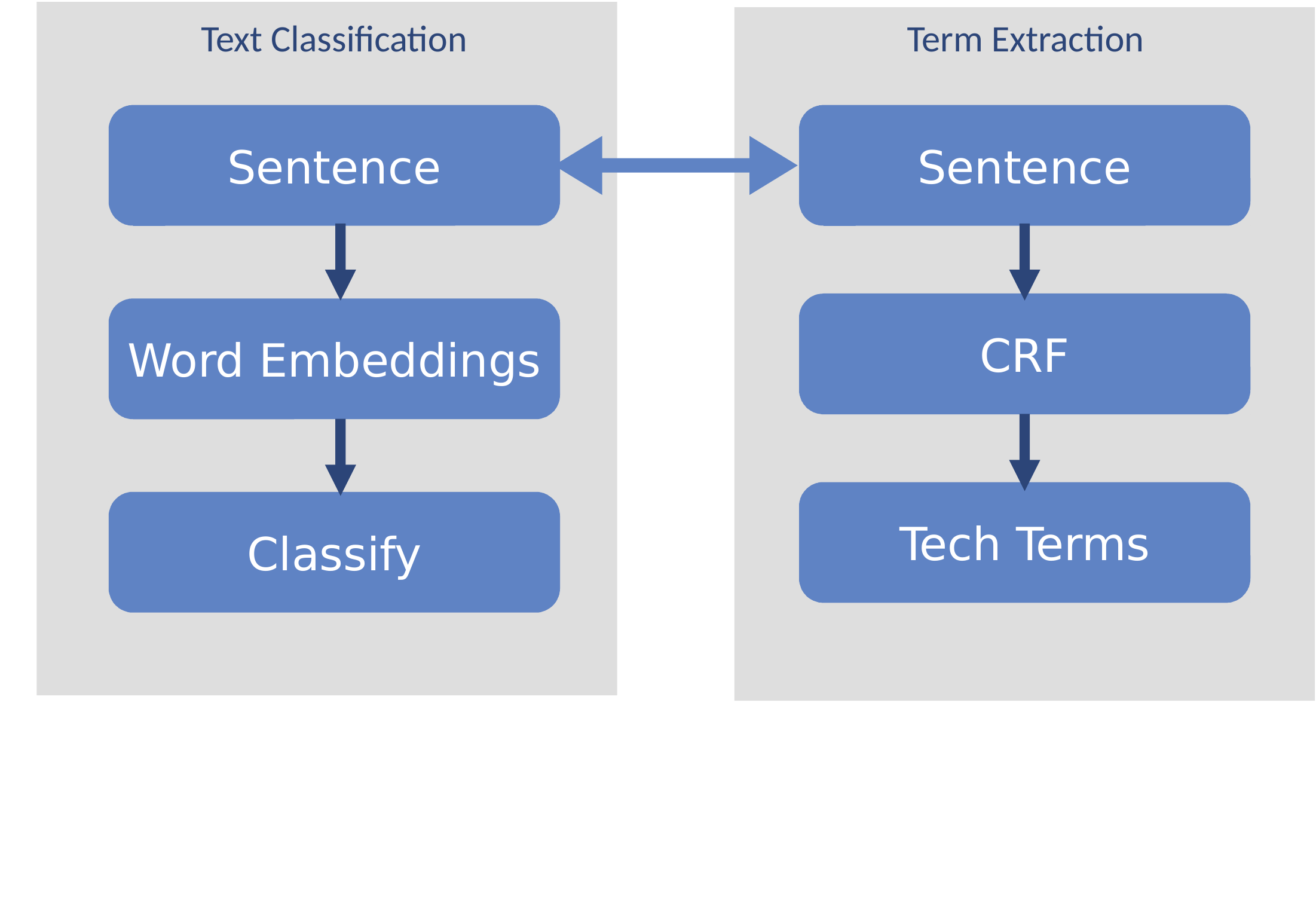}
\caption{Detailed system diagram of TEST.}
\label{fig:TEST-system}
\vspace{-0.2cm}
\end{figure}

In summary, our TEST system can be schematically described as Fig.~\ref{fig:TEST-system}. The TEST system is based on two-stage cascading model -- text classification and term extraction. 

\section{Results and Evaluation}
\label{sec:results}
We evaluated the performance of our proposed system based on the performance of each of the two stages of the cascading method. As discussed earlier, the first stage deals with the sentence classification, whereas the second stage deals with the term extraction. 

\subsection{Subjective Evaluation}

\begin{figure}[htb]
\centering
\subfloat[Sample illustration shows how sentence level and term level detection can catch different technology terms.]{\includegraphics[width=0.45\textwidth]{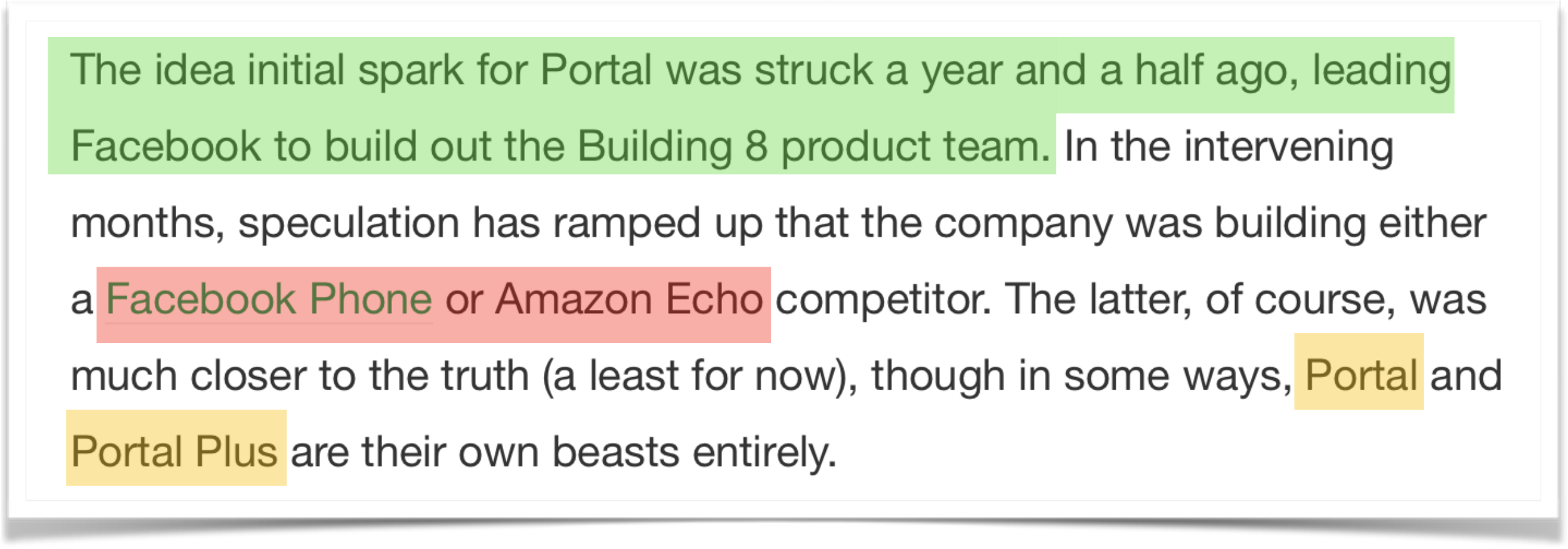}}\\
%\vspace{-4mm}
\subfloat[Sample illustration shows that the sentence level and term level detection can overlap in a few cases. ]{\includegraphics[width=0.45\textwidth]{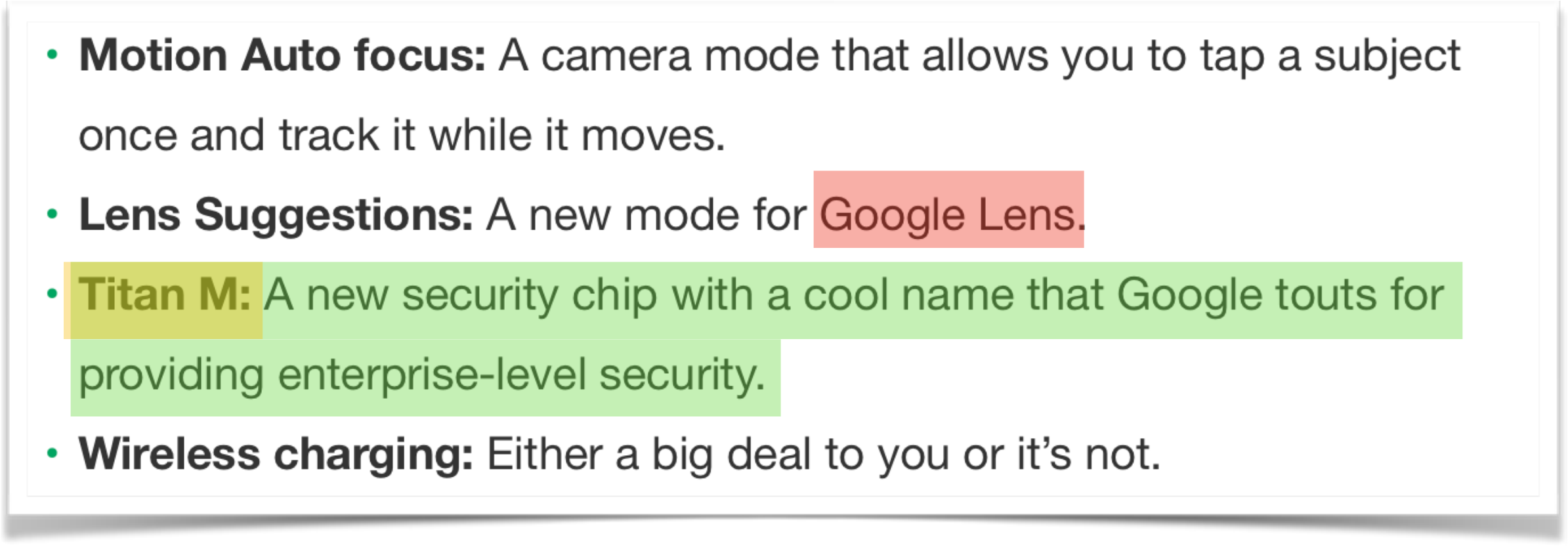}}\\
\subfloat[Sample illustration demonstrates that the system can detect multiple sentences and technology terms.]{\includegraphics[width=0.45\textwidth]{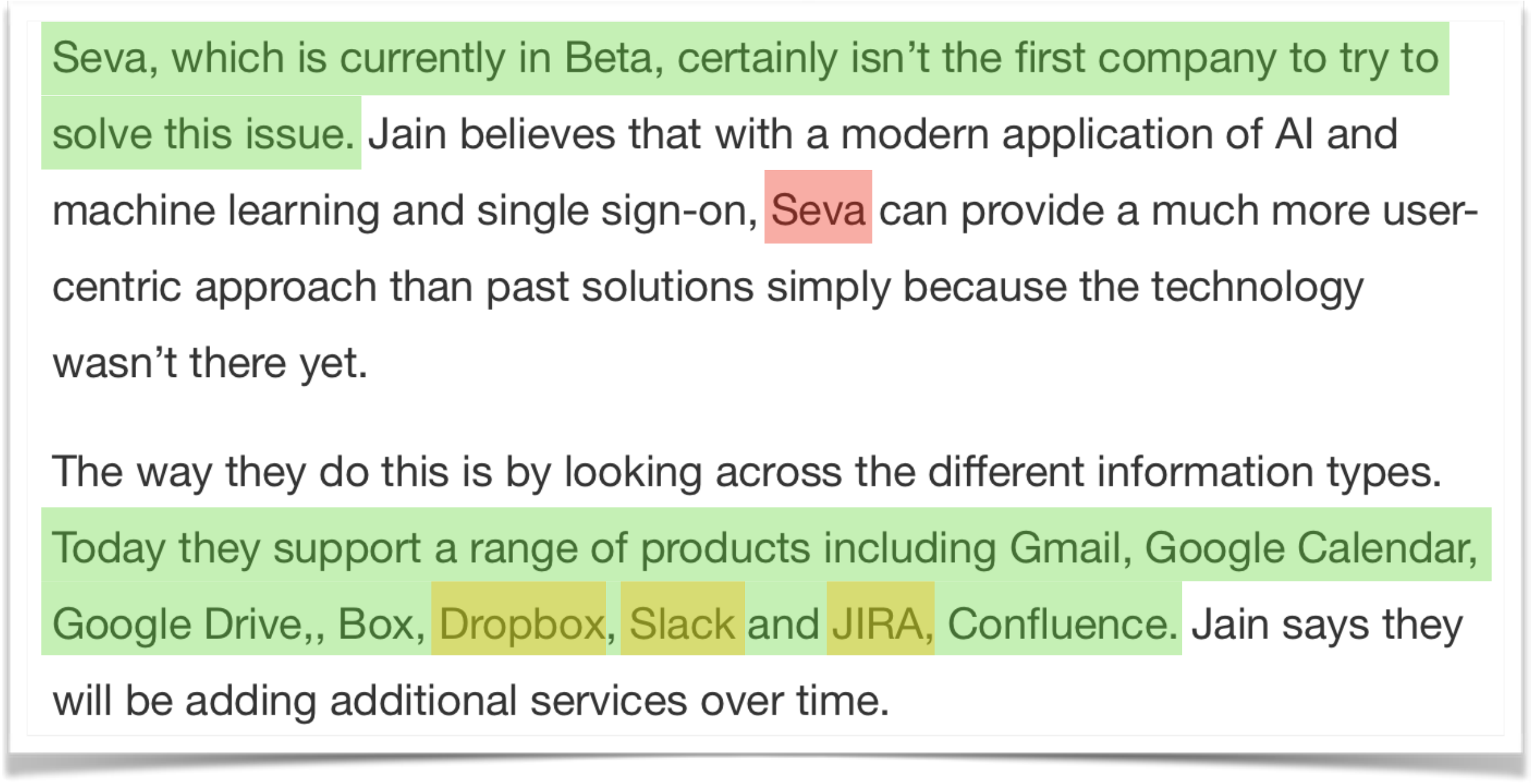}}
\caption{Illustration showing the results in executing the TEST system against technology articles. The green colour marks the sentences with potential technology terms which were detected by the system. The yellow colour marks the technology terms that are detected by the system. Finally, the red colour marks the terms that are missed by the system.}
\label{fig:few-examples}
\end{figure}

In Fig.~\ref{fig:few-examples}, we present a few examples of how our system performs at run-time. In these examples, our proposed TEST system examines the text in certain technology articles found on the web. The system returns a list of sentences with potential technology terms, and another list of technology terms (can be multi-word technology term). 
In Fig.~\ref{fig:few-examples}(a), our proposed system identifies that the following sentence \emph{The idea initial spark for Portal \ldots production team.} contains a technology word. It also identifies the technology terms such as \emph{Portal} and \emph{Portal Plus} as indicated in yellow. However, in some cases the system misses some technology terms, such as \emph{Facebook Phone} or \emph{Amazon Echo}. Our system in these cases gets confused when organisation names make part of the technology term. Hence, it displays an incorrect prediction. Similar observations can be found in Fig.~\ref{fig:few-examples}(b) and Fig.~\ref{fig:few-examples}(c).

\subsection{Objective Evaluation}
In order to provide an objective evaluation of our system, we report the F-score of the two stages. We use F-score measure to evaluate how good our system performs in a real-world dataset.

Suppose, $TP$, $FP$, $TN$ and $FN$ represent the true positive, false positive, true negative and false negative samples of the task. The F-score is defined as:

\begin{equation}
\label{eq:fs}
\mbox{F-Score} = \frac{(2\times\mbox{Precision}\times\mbox{Recall})}{(\mbox{Precision}+\mbox{Recall})},
\end{equation}

where precision and recall are respectively defined as $\mbox{Precision} = \frac{TP}{TP+FP}$ and $\mbox{Recall} = \frac{TP}{TP+FN}$.

In the first stage comprising the sentence classification part, we split our dataset of $20$ thousand sentences into $3$ parts -- training set $70\%$, validation set $15\%$ and test set $15\%$. We maintained a balanced distribution between the positive and negative examples while splitting. In the second stage comprising the term extraction from a sentence, we used a similar split. We ensured that every token in the data gets a positive (technology) or a negative (non technology) tag. We also used the same assessment metric of F-score to evaluate how well this stage performs. 

Table~\ref{table:result} summaries the performance of our proposed system. We observe that the proposed TEST system has a competitive score in both its stages. We obtain a F-score of $0.93$ and $0.96$ respectively in both the stages. 

\begin{table}[htb]
%\normalsize
\centering
\begin{tabular}{c||c|c}
  \textbf{Stage}  & \textbf{Activity}  &  \textbf{F-score}  \\
  \hline
  I & Sentence classification & 0.93 \\     
  II & Term extraction & 0.96\\
\end{tabular}
\caption{Performance evaluation of our proposed system.}
\label{table:result}
\vspace{-0.5cm}
\end{table}

It is important to benchmark these results with other similar technology text extraction systems. However, because of the lack of the existence of such similar systems, we could not benchmark our performance with other similar systems.

\section{Conclusion}
\label{sec:conclusion}
In this paper, we propose an end-to-end system called TEST that automatically extracts the technology terms from text. Our system is trained on a large corpus of technology-related news articles and blogs, and has a competitive performance in detecting and extracting tech terms from a sentence. In the future, we plan to use the sentiment around these technologies in order to evaluate and predict the impact of those technologies and tools in any specific AI area. In addition to word embeddings, we also plan to borrow techniques from convolutional neural networks (CNN), Recurrent Neural Networks (RNNs), Long Short Term Memory Units (LSTMs) etc., to propose an ensemble-based TEST system.

\balance

\begin{acks}
The ADAPT Centre for Digital Content Technology is funded under the SFI Research Centres Programme (Grant 13/RC/2106) and is co-funded under the European Regional Development Fund.

\end{acks}

\bibliographystyle{ACM-Reference-Format}

\end{document}